\documentclass{vgtc}                          

\graphicspath{{figures/}{pictures/}{images/}{./}} 
\usepackage{enumitem}
\usepackage{times}                     

\usepackage{tabu}                      
\usepackage{booktabs}                  
\usepackage{mwe}                       
\usepackage{todonotes}
\usepackage{tabularx}
\usepackage{mathptmx}                  

\onlineid{0}

\vgtccategory{Research}

\vgtcinsertpkg

\title{
Human Factors in Immersive Analytics 

}

\author{Yi Li\thanks{e-mail: yi.5.li@tuwien.ac.at}\\ %
        \scriptsize TU Wien %
\and Kadek Satriadi\thanks{e-mail: kadek.satriadi@monash.edu}\\ %
     \scriptsize Monash University %
\and Jiazhou Liu\thanks{e-mail: joe.liu@monash.edu}\\ %
     \scriptsize Monash University %
\and Anjali Khurana\thanks{e-mail: anjali\_khurana@sfu.ca}\\
    \scriptsize Simon Fraser University %
\and Zhiqing Wu\thanks{e-mail: zwu755@connect.hkust-gz.edu.cn}\\
    \scriptsize HKUST (Guangzhou) %
\and Benjamin Tag\thanks{e-mail: benjamin.tag@unsw.edu.au}\\
    \scriptsize University of New South Wales %
\and Tim Dwyer\thanks{e-mail: tim.dwyer@monash.ed}\\
    \scriptsize Monash University}

\abstract{
It has been ten years since the term ``Immersive Analytics'' (IA) was coined and research interest in the topic remains strong. Researchers in this field have produced practical and conceptual knowledge concerning the use of emerging immersive spatial display and interaction technologies for sense-making tasks through a number of papers, surveys, and books. However, a lack of truly physically and psychologically ergonomic techniques, as well as standardized human-centric validation protocols for these, remains a significant barrier to wider acceptance of practical IA systems in ubiquitous applications. Building upon a series of workshops on immersive analytics at various conferences, this workshop aims to explore new approaches and establish standard practices for evaluating immersive analytics systems from a \textit{human factors} perspective. We will gather immersive analytics researchers and practitioners to look closely at these human factors---including cognitive and physical functions as well as behaviour and performance---to see how they inform the design and deployment of immersive analytics techniques and applications and to inform future research. 

} 


\begin{document}


\firstsection{Introduction}
\maketitle
\begin{figure}
    \centering
    \includegraphics[width=\linewidth]{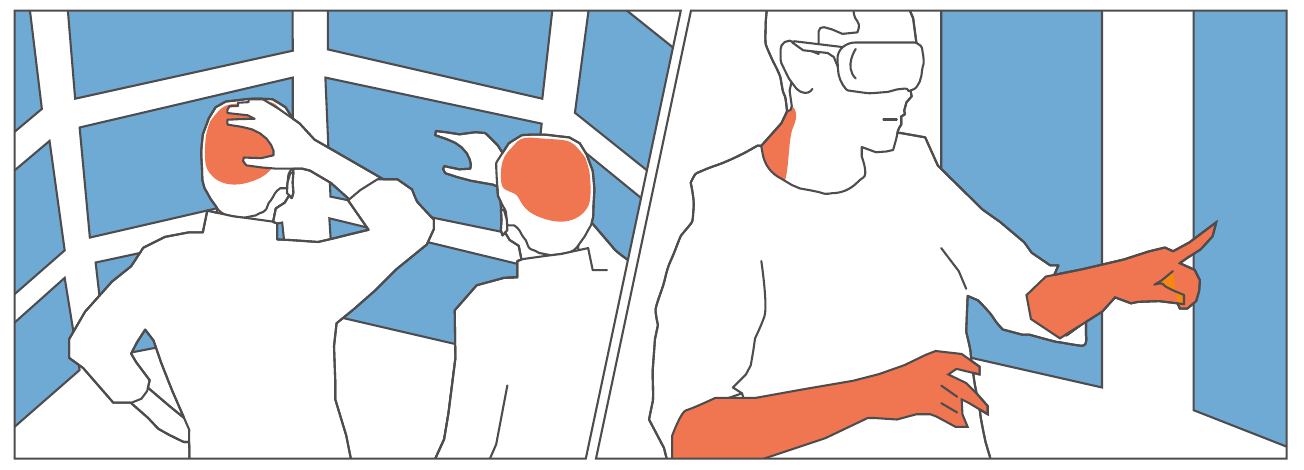}  \caption{This workshop will look closely at mental (illustrated as the red parts of the users in the left figure) and physical aspects (red parts in the right figure) involved in immersive analytics applications, including how to better measure related factors.}
  \label{fig:teaser}
\end{figure}


The field of Immersive Analytics (IA) is inspired by the idea that work environments that bring data into the space around the user (or \textit{users} in collaborative contexts) promote effective and engaging sensemaking activities~\cite{dwyer2018immersive}. Steady development and release of new devices---such as Apple's Vision Pro, Meta headsets (e.g., Quest 3 and 3s) and glasses (e.g., Orion glasses), Pico's Pico 4 Ultra, Varjo's XR 4, and more---demonstrates a positive trend for the future of IA, with technical barriers to adoption gradually dropping away. Yet, our understanding of the human aspects of IA---i.e., the capabilities of humans to operate with information presented in immersive environments---lags behind the rapid development of the supporting hardware. 

Human factors---including cognitive and physical functions, behaviour, and performance---in IA applications still pose open research questions. So far, we have witnessed significant milestones in the field, including---but not limited to---a textbook~\cite{marriott2018immersive}, surveys of IA and situated analytics (IA employing Augmented Reality to situate information in a real-world evironment)~\cite{fonnet2019survey, zhao2022metaverse, kraus2022immersive, saffo2023unraveling, Lee:2023:DesignPatterns, Satriadi:2023:ProxSituated, bressa2021s}, lists of challenges and agendas\cite{skarbez2019immersive, ens2021grand}, and results on graphical perception\cite{whitlock2020graphical}. Nevertheless, we have yet to start exploring the theoretical aspects of IA, as shown by a recent review~\cite{saffo2023unraveling} and mentioned in a grand challenge paper~\cite{ens2021grand}. It is apparent that the ``what technique works'' topics have been seen more often than ``why a technique works'', especially when we reflect on the role of cognitive and physical factors of the user. Considering the growing interest in measuring IA techniques beyond performance, such as memory \cite{liu2022effects, liu:2024:LandmarkSpatialMemory, han2023evaluating, hurter2024memory}, learning~\cite{khurana2024just}, physical reach and strength~\cite{wu2024toward, li2023revisiting}, and physical fatigue~\cite{dai2024precise, li2024nicer,li2025alphapig}, it is the right time to define relevant human factors in IA, and how we should go about measuring them.  

This workshop aims to contribute to this landscape by defining a roadmap for new directions in understanding the human characteristics when interacting with immersive content and therefore to optimize the design of IA systems as well as identify new opportunities and applications. Through a combination of synchronous and asynchronous sessions, along with in-person discussion, we will gather researchers from diverse backgrounds and expertise. Participants will come from IA and related fields, including interaction design, data visualization, VR, AR, biomechanics, embodied cognition, and more, offering a broad array of viewpoints.

By fostering interdisciplinary collaboration of experts, this workshop will pave the way for defining key human factors and establish robust methodologies for measuring and optimizing user interaction with immersive analytics systems.

\section{Previous Workshops}
This workshop will build upon topics discussed in previous workshops on immersive analytics, as listed below.

\begin{enumerate}[leftmargin=*,itemsep=1pt, topsep=2pt, parsep=0pt, partopsep=0pt]
    \item ACM ISS 2016~\cite{bach2016immersive}, \textit{Immersive Analytics: Exploring Future
Interaction and Visualization
Technologies for Data Analytics}. Real-world visual analytics, collaboration, hybrid 2D and 3D, interaction affordances, changing technologies, application areas, platforms and toolkits. 
    
    \item IEEE VIS 2017~\cite{bach2017immersive}, \textit{Immersive Analytics: Exploring Future Visualization and Interaction
Technologies for Data Analytics}. Changing technology and scenarios, hybrid 2D and 3D, affordances for immersion, collaboration, physical and tangible visualization, interaction techniques, real-world IA and applications.

\item ACM CHI 2019~\cite{bach2019interaction}, \textit{Interaction Design and Prototyping for Immersive Analytics.} 
Focus on prototyping and design sessions.

\item ACM CHI 2020~\cite{ens2020envisioning}, \textit{Envisioning Future Productivity for
Immersive Analytics}. Productive user interfaces for IA (e.g., user productivity, multimodal input, collaborative analytics, users and scenarios, situated and spatial analytics, evaluation).

\item ACM ISS 2022~\cite{ISSImmersiveAnalytics2022}, \textit{Workshop on Immersive Analytics Spaces and Surfaces.}  Exploring benefits and applications of flat versus spatial interaction affordances within immersive environments. 

\item ACM CHI 2022~\cite{ens2022immersive}, \textit{Immersive Analytics 2.0: Spatial and Embodied Sensemaking}. The workshop focused on understanding immersive versus non immersive platforms for sensemaking and analytics. Topics include fundamental research questions, new evaluation methods, human-AI sensemaking, accessibile sensemaking, data physicalization, situated/embedded data visualization, and authoring immersive visualization.

\end{enumerate}

Each workshop iteration uncovers fresh opportunities in the rapidly evolving technological and research landscape. 
Now, three years post-CHI 2022, we expect significant innovations focused on human factors challenges. While psychology and philosophical theories of embodied interaction have been discussed in previous workshops as motivation for design ideas, they have not been deeply explored. 

\begin{table*}
\small
\centering
\renewcommand{\arraystretch}{1.4} 
\begin{tabularx}{\linewidth}{|>{\bfseries}p{2.8cm}|l|X|}
\hline
\textbf{Activity} & \textbf{Duration} & \textbf{Description} \\ \hline
Welcome & 5 minutes & Organisers introduce themselves, provide workshop motivation, and overview of activities. \\ \hline
Keynote & 30 minutes + 15 minutes Q\&A & Domain expert keynote to set the tone with challenging questions, engaging stories, and examples. Time for discussion included. \\ \hline
Group Discussion 1 & 30 minutes & Participants split into groups (5-8) to focus on key human factors: Cognitive, Perceptual, Biomechanical, Sensory-Motor, and Collaboration. \\ \hline
\textit{Break} & \textit{40 minutes} & \textit{Lunch Break.} \\ \hline
Video Showcase & 30 minutes & Participants showcase relevant work through videos and demos, followed by an informal session to engage with presenters. \\ \hline
Group Discussion 2 & 40 minutes & A second round of group discussions, rotating focus on human factors. Groups formulate a research agenda with challenges, opportunities, and goals. Concludes with lightning presentations. \\ \hline
Closing & 10 minutes & Review of activities, summary of takeaways, group discussion on next steps, future workshops, and potential publications. \\ \hline
\end{tabularx}
\caption{Proposed Schedule}
\label{tab:schedule}
\end{table*}

\section{Workshop Aims}
This workshop will establish a research agenda for broadening contributions made so far in IA, especially on human capabilities and limitations independent of specific technology.
One of the core workshop outcomes will be a systematic review paper that outlines key research questions, identifies new application areas, and proposes better of evaluation methods.
Additional aims are:
\begin{itemize}[leftmargin=*,itemsep=1pt, topsep=2pt, parsep=0pt, partopsep=0pt]
    \item Identification of fundamental human factors that contribute to or limit the effectiveness of IA.
    \item Broadening our understanding of physical and cognitive benefits and limitations of IA to inform an improved expansion of IA into further application areas.
    \item To identify approaches for evaluating these human factors and techniques for addressing their limitations.
\end{itemize}




\section{Perspectives of Interest}
\label{sec:challenges}
Research questions will focus on cognitive and physical aspects: 

\noindent\textbf{Fundamental Cognitive Factors In System Design} -- 
Understanding the roles of cognition and perception---including attention, memory, knowledge, visual imagery, language, problem solving, reasoning and decision making, and relevant scenarios---crucial for improving IA interface design. 

    

\noindent\textbf{Fundamental Physical Factors In System Design} -- Using body movements as input to interact with IA systems is widely adopted in literature. However, little discussion has yet been made about how these input modalities support analytical tasks. Factors that require further consideration include human biomechanical limitations on movement such as maximum speed, accuracy and fatigue over time.

\noindent\textbf{Synergistic Effects Between Cognitive And Physical Factors} -- Novel gestures were proposed to manipulate diverse types of data and information. We will discuss how users' mental models of such techniques influence their comprehension and decision-making processes in analytic tasks with consideration of theories of distributed cognition and kinesthetic memory effects.

\noindent\textbf{New Evaluation Methods And Research Methods} -- Previous system evaluation relies on performance metrics, such as, completion time and accuracy, and post-study subjective preference. New research agendas are needed to establish a better practice of system evaluation using recently developed techniques for measuring cognitive and physical costs during interaction. 

\noindent\textbf{Novel Interaction Techniques} -- Cognitive and physical costs are often mentioned when IA techniques are proposed. However, specifically how these solutions can enhance users' overall data interpretation, insight generation, and decision-making processes and in what type of analytic tasks remains unsolved.

\noindent\textbf{Integration with AI} -- Emerging AI model capabilities for real-time data processing, predictive modelling, and intelligent pattern recognition offer real-time and in-situ guidance in task-embedded scenarios. In such human-AI interactions, humans are not just passive data recipients; they actively interpret, question, and act on information. Integrating AI with IA could enhance situational awareness by enabling more intuitive and context-rich interactions with complex data environments. However, cognitive load, perception, attention, and decision-making processes must be carefully considered. AI must support—--not overwhelm—--users by presenting data in a way that aligns with natural human cognition and situational needs. 

\noindent\textbf{Adaptive User Interfaces} -- Mathematical optimization is often used to adapt interfaces for visibility and readability. Recent toolkits facilitate optimization and abstract the formulation of such mathematical problems. Novel formulations of cognitive and physical factors as adaptation objectives and constraints are required to make optimization accessible to creators. Additionally, IA applications have unique requirements (e.g. 3D and environmental constraints) that must be identified and addressed.

\section{Organisers}
We are academics and representatives from industry with diverse backgrounds (i.e., IA, V/A/XR interaction, empathic computing, and affective computing). We include organizers from previous IA workshops (Satriadi, Liu,
Dwyer). Our diversity includes balanced female and male, senior, mid-career, and early-career academics, as well as students with expertise in accessible and inclusive VR, and mental models in AR interactions. We represent four different continents and six different institutions.

\noindent\textbf{Yi Li}, Postdoctoral Fellow, Artifact-based Computing and User Research Unit, TU Wien, Austria. Her focus is mitigating physical fatigue in adaptive XR interaction. She recently published a comprehensive shoulder fatigue model for mid-air interaction in SIGGRAPH 2024~\cite{li2024nicer} and a meta technique to develop fatigue-aware XR interaction~\cite{li2025alphapig} in CHI 2025.

\noindent\textbf{Kadek Satriadi} is a Lecturer at Monash University Australia in the Embodied Visualisation research group with research on IA for geospatial data, situated scenarios, storytelling, and physicalization. He co-organised IA workshops at ISS 2022 and CHI 2022 and was a technology co-chair at IEEE VIS 2023.  

\noindent\textbf{Jiazhou `Joe' Liu}, Postdoctoral Fellow, Monash EmVis group. He co-organised the CHI2022 ``IA 2.0'' workshop. His foci are immersive visualisation view management, spatial cognition and skills. His recent research explores AI in IA, with applications in medical and construction domains.

\noindent\textbf{Anjali Khurana}, Ph.D. student, Simon Fraser University (SFU), Canada. She seeks to understand users’ mental models in XR gestural interaction and interactions with LLMs. She aims to invent onboarding and assistance techniques novices in these technologies. She has published in ACM CHI 2025~\cite{khurana2025me}, ACM IUI 2024~\cite{khurana2024and}, and ACM IMWUT 2023~\cite{khurana2024just}.

\noindent\textbf{Zhiqing Wu}, PhD student, Hong Kong University of Science and Technology (Guangzhou), China. Her aims to make VR interaction more inclusive and accessible. Her CHI2024 paper was on understanding VR interaction for older adults.


\noindent\textbf{Benjamin Tag}, Sen.\ Lecturer, UNSW Sydney. HCI, Affective Computing, and Human-AI Interaction. His research focuses on technologies for assessing mental~\cite{Tag2019} and physical states~\cite{li2024nicer} in real-world and virtual environments, emphasising human emotion, cognitive psychology, and context-aware computing. 

\noindent\textbf{Tim Dwyer}, Professor of Computer Science, EmVis Group, Monash, Australia. He has explored topics in immersive data visualisation for 24 years and co-authored the earliest papers establishing IA (e.g. \cite{7314296}) and edited the IA book \cite{marriott2018immersive} after co-organizing the first IA Dagstuhl Seminar.

\section{PRE-WORKSHOP PLANS}
The timeline for the workshop organization is as follows:
\begin{center}
\begin{tabular}{ll}
\multicolumn{1}{l}{\textbf{Activity}} & \multicolumn{1}{l}{\textbf{Date}} \\ \hline
Website and CFP launches                & May 15, 2025                  \\
EOI deadline                            & Aug 1, 2025                   \\
Acceptance notification                 & Aug 5, 2025                      \\
Asynchronous material shared            & Oct 1, 2025                         \\
Workshop at VIS 2025                    & Nov 1, 2025                          \\ \hline

\end{tabular}
    
\end{center}

We will call for participation via a dedicated website (\autoref{sec:CFP}), the ACM
CHI and IEEE VIS mailing lists, as well as the organizer's networks and social media. Our website will provide participants background and motivation and general instructions on pre-workshop activities, submission deadlines, and organiser contacts. In the later stage, workshop submissions (e.g., written statements and/or videos) may also be made publicly available on the website upon acceptance. 

\subsection{Workshop Structure}
The workshop will be run \textbf{in-person} to ensure a high quality of engagement for participants. We aim to attract an in-person workshop attendance limit of 35. If we receive more than 35 applications, then we will select participants based on the quality and suitability of their submissions using a lightweight peer-review process, selecting reviewers from the wider community as necessary. We will publish the workshop proceedings on CEUR-WS.org (\url{https://ceur-ws.org/}). We will focus on ideation through discussion; hence, we do not expect to have paper presentations, but will follow up with a publication plan and future plans (see section \ref{post}). We will accommodate accessibility for special requirements as requested by participants in the registration process.

\section{ASYNCHRONOUS ENGAGEMENT}
We will share a Google Drive folder containing all related activities prior to the workshop. These include a workshop brief, schedule, and code of conduct. Prior to the workshop, participants will be asked to prepare a short written 
and/or video response addressing our proposed research topics and challenges (\autoref{sec:challenges}). These may include examples from their own work or other research related to their expertise. The submitted responses will be stored in a secured Google Drive folder and reviewed by the workshop organizers. This activity will be conducted asynchronously. 

\section{MAIN WORKSHOP ACTIVITIES}
To prompt group discussions we will feature lightning talks and live demonstrations. The schedule format \autoref{tab:schedule} is informed by previous successful VIS workshops (i.e., two 75-minute sessions for a \textbf{half-day} workshop).

\section{SUCCESSFUL OUTPUTS AND FOLLOW-UP}
\label{post}
The success of the workshop will be measured by follow-up outputs, activities and community building, especially:
\begin{itemize}[leftmargin=*,itemsep=1pt, topsep=2pt, parsep=0pt, partopsep=0pt]
\item A formulation of the recent innovations, research challenges, short-term and long-term directions of IA research grouped by the topics covered in Section \ref{sec:challenges}. This formulation will guide our future workshop to be submitted to ACM CHI 2026.
\item New and stronger collegial networks among AI researchers, particularly in the topics of cognitive and physical aspects. This network is intended to lead to research and grant collaborations.
\end{itemize}

\noindent\textbf{Concrete Outcomes:}
\begin{enumerate}[leftmargin=*,itemsep=1pt, topsep=2pt, parsep=0pt, partopsep=0pt]
    \item \textbf{Position paper} -- a peer-reviewed position paper including a comprehensive report on the state of understanding of human factors in IA and an agenda for future research.
    \item \textbf{Working group papers} -- focused discussion groups will spawn new networks leading to independent research and eventually co-authorship on these specific sub-topics.
    \item \textbf{Future workshops} -- more specialized subtopics within human factors in IA as well as application to future effective IA system design and technological as well as societal readiness.
\end{enumerate}


\section{CALL FOR PARTICIPATION}
\label{sec:CFP}
Join us for an exploration of topics on ``Human Factors in Immersive Analytics'' workshop, to be held at IEEE VIS 2025. This workshop will focus on the cognitive, perceptual, biomechanical, sensory-motor, and collaborative aspects of Immersive Analytics and how these factors influence sensemaking and data comprehension. While immersive technologies have shown potential in analytics, many open research questions remain concerning the human factors that shape the effective use of immersive analytics applications. 
We encourage researchers, practitioners, and designers to submit research papers, early-stage work, or conceptual ideas, ranging from 1 to 4 pages (excluding references), in the standard TVCG submission format. 
Each paper will be reviewed by the organizers to ensure quality and relevance. 
All accepted papers will be compiled and made available to all participants ahead of the workshop, facilitating pre-event engagement and discussion. A selection of papers will be published after the workshop on the CEUR-WS.org website, where authors can choose to opt in for it. At least one author of each accepted submission must register and attend the workshop. The program will consist of a keynote, live demonstrations, and group discussions, focusing on topics such as fundamental cognitive factors in system design, fundamental physical factors in system design, the synergistic effect between cognitive and physical factors, new evaluation and research methods, novel interaction techniques, and adaptive user interfaces. More details about the workshop can be found at \url{https://sites.google.com/monash.edu/iahf-workshop}. 

\bibliographystyle{abbrv-doi}

\bibliography{sections/bibliography}
\end{document}